\documentclass[preprint]{aastex}

\def\lesssim{\mathrel{\hbox{\rlap{\hbox{\lower4pt\hbox{$\sim$}}}\hbox{$<$}}}}
\def\gtrsim{\mathrel{\hbox{\rlap{\hbox{\lower4pt\hbox{$\sim$}}}\hbox{$>$}}}}

\begin{document}

\title{Where is the Inner Edge of an Accretion Disk Around a Black Hole?}

\author{Julian H. Krolik}
\affil{Physics and Astronomy Department, Johns Hopkins University,
    Baltimore, MD 21218}
    
\author{John F. Hawley}
\affil{Department of Astronomy, University of Virginia,
Charlottesville VA 22903}
\shorttitle{Inner Edges Inner Edges}

\begin{abstract}

    What is meant by the ``inner edge" of an accretion
disk around a black hole depends on the property that defines the edge.
We discuss four such definitions using data from recent high-resolution
numerical simulations.  These are: the ``turbulence edge", where
flux-freezing becomes more important than turbulence in determining the
magnetic field structure; the ``stress edge", where plunging matter
loses dynamical contact with the outer accretion flow; the ``reflection
edge", the smallest radius capable of producing significant X-ray
reflection features; and the ``radiation edge", the innermost place
from which significant luminosity emerges.  All these edges are
dependent on the accretion rate and are non-axisymmetric and
time-variable.  Although all are generally located in the vicinity of
the marginally stable orbit, significant displacements can occur,
and data interpretations placing the disk edge precisely
at this point can be misleading.  If observations are to be used successfully
as diagnostics of accretion in strong gravity, the models used to interpret
them must take careful account of these distinctions.

\end{abstract}


\section{Introduction}

    Accretion disks around a star end at the star's surface; the inner
radius of a disk around a black hole is less well determined.  Because
black holes have no hard surface, more complicated dynamical processes
define what is meant by a disk's ``inner edge."  The best-known of these
arises from general relativity itself, which forbids stable circular
orbits inside the critical radius of the marginally stable orbit,
$r_{ms}$, not far outside the event horizon.  

In a standard thin accretion disk, gas follows very nearly
circular orbits and drifts inward because a nonzero stress removes
angular momentum from fluid elements and transfers it outward.  Inside
$r_{ms}$, fluid elements can spiral into the hole on free-fall orbits,
i.e., without further loss of angular momentum or energy.  In many
treatments of black hole accretion disks, it is therefore explicitly (or
tacitly) assumed that the inner edge of the disk is at, or at least
very close to, $r_{ms}$.  Although there is clearly a dynamical
transition from rotational support to free-fall that must occur in the
vicinity of $r_{ms}$, whether this constitutes the disk inner boundary
depends on precisely what one means by that term. The location
of the inner edge relative to $r_{ms}$ depends on which property one is
examining, and departures from $r_{ms}$ can, in many instances, be
significant.  Dynamical issues require a different definition from
observational characteristics.

    For example, one definition of inner boundary is the
innermost radius from which significant luminosity emerges to the
outside world, the {\it radiation edge}.  The inner regions of disks
may grow dim for a variety of reasons: a decrease in the radiative
efficiency due to flow dynamics, gravitational redshift and other
general relativistic effects, cessation of dissipative heating (as in
Page \& Thorne 1974), or photon trapping (an effect particularly strong
in ``slim disks", Abramowicz et al. 1988).  All of these effects
taken together determine precisely where the disk ceases to radiate.

    As another example, even if little radiation originates from gas
near the hole, there may still be enough material inside $r_{ms}$ to
reflect and reprocess incident X-ray photons.  Reynolds \& Begelman
(1997) suggested that this effect may explain the shape of the Fe
K$\alpha$ line in MCG--6-30-15.  To what extent this is possible
depends on details of the accretion flow (e.g., Young, Ross \& Fabian
1998).  Thus, for any particular accretion state, one may define a {\it
reflection edge} that may be well separated from the marginally stable
orbit---or from the radiation inner edge.  This distinction is an
important one: for example, the assumption that the reflection edge
is identical to $r_{ms}$ has been used as the basis for determining
the spin parameter of a black hole (Wilms et al. 2002).

    Other inner edge definitions are based not on radiative properties,
but on the underlying dynamics.  The {\it stress edge} can be defined
as the point where accreting matter loses dynamical contact with the
disk it left behind.  In a purely hydrodynamic model using the
traditional Shakura-Sunyaev $\alpha$ stress proportional to the
pressure, this edge would be quite close to $r_{ms}$ because of the
rapid drop in pressure as the gas accelerates inward (e.g., Abramowicz
\& Kato 1989).  Our growing confidence that magnetic stresses account
for angular momentum transfer within disks (Balbus \& Hawley 1998),
however, has modified this view.  Magnetic stress can continue well
inside $r_{ms}$ (Krolik 1999a; Gammie 1999; Hawley \& Krolik 2001,
2002, hereafter HK01, HK02; Reynolds \& Armitage 2001), so $r_{ms}$ may
not be a good estimate of the stress edge.

    Yet another inner edge, which should generally be outside the
stress edge, is the {\it turbulence edge}.  This is the place
where the magnetic field switches from being controlled by the
mechanics of saturated magnetohydrodynamic (MHD) turbulence to
simple flux-freezing.  Here the internal disk structure begins to
change, as the gas prepares for its final plunge.

  Although we speak of these as edges, we emphasize that none of
them is sharp.  At any given instant the scale lengths of the
transitions in question can be comparable to the radius.  Further,
because the flows are very strongly time-variable, instantaneous edge
positions can change substantially from time to time.  Consequently,
all of them should be thought of in terms of zones within which the
relevant transition can occur.
  
     In this paper we use data from recent black hole accretion
simulations for a dual purpose.  On the one hand, we
examine and clarify the quantitative relations between
the different inner edge definitions.  Our aim is to create a
conceptual framework and an associated language for discussing
these issues.  As our understanding of accretion dynamics
deepens, we expect these distinctions to become more prominent.

     On the other hand, we also develop several observational
consequences of these distinctions.  As we enter an era in which
detailed models are fit to a variety of relativistically-shifted
and broadened features, it is important to clearly define the radial
emissivity distributions governing these features.  Precision on these
issues is vital when observations are to be used as direct tests of
specific general relativistic properties of black holes.  Although
the numerical values we present for some of these quantities are
uncertain because of the approximations made in current simulations,
data from future simulations, when employed in the framework
defined here, should enable finer definition of these edges and
therefore more reliable inferences.

\section{The Turbulence Edge}

      We begin with the disk edges that are defined in terms of dynamical
properties.  Disk dynamics are governed by MHD
turbulence and the resulting Maxwell stresses.  Within the disk, the
intensity and structure of the magnetic field are determined by a
balance between the underlying magneto-rotational instability, or MRI
(Balbus \& Hawley 1991), which generates the field, and the field loss
terms, including local resistive dissipation of short length-scale
fluctuations, and upward motion of buoyant magnetic flux (as studied,
e.g., by Miller \& Stone 2000).  At large radius the timescales for
these processes are short compared to the inflow time, but close
to $r_{ms}$ the time required for energy to travel down the
turbulent cascade becomes longer than the inflow time, and the magnetic
field instead evolves by flux-freezing.   We define the {\it turbulence edge}
as the boundary between the region where the magnetic field
dynamics are dominated by the turbulent cascade and the region
controlled by flux-freezing.

     The location of this edge may be estimated in any of several ways.
We can compare the timescales for the competing processes, or we can look for
a change in the field structure due to this transition, or we can compare
the magnitude of the turbulent velocities to the magnitude of the inflow
speed.  We will take up each of these in turn.

\subsection{Timescales}

    Consider the timescale approach first.  In a time-steady
disk with accretion rate $\dot M$, the azimuthally- and vertically-integrated
energy per unit area available for dissipation at (cylindrical) radius
$r$ is the difference between the net deposition of potential energy by
accretion and the net work done by inter-ring torques:
\begin{equation}
Q = {-\dot M \over 2\pi r}{d \over dr}
        \left[ E_B + \Omega \left(j - j_{in}\right)\right],
\end{equation}
where $E_B$ is the binding energy per unit mass, $j=r^2\Omega$ is the angular
momentum per unit mass, $j_{in}$ is the specific angular momentum
of matter accreted by the black hole, and $\Omega$ is the orbital
frequency.  To be precise, relativistic corrections should be
applied to these quantities (Novikov \& Thorne 1973, Page \& Thorne
1974), but the Newtonian formalism expresses
the physics more transparently.

      If the torque is largely due to MHD turbulence in which the
radial and azimuthal magnetic field components are correlated, this
energy is initially given primarily to the field and secondarily to
turbulent motions of the matter.  In the simulations of HK02,
for example, within the disk proper (i.e., well outside
$r_{ms}$) the local ratio of magnetic field to random motion energy
density varies from $\sim 1$ to $\sim 100$.  This energy is then
transferred from relatively long lengthscale motions to shorter
lengthscale motions in a turbulent cascade whose eddy turnover time (in
the same simulations just cited) is typically comparable to an orbital
period.  By definition, the lengthscale at which the cascade cuts off
is the lengthscale on which the dissipation rate becomes as fast as the
nonlinear energy transfer rate.  Under ordinary conditions, this may be
many orders of magnitude smaller than the stirring scale.
Consequently, the total time required to traverse the cascade would
then be a logarithmic factor, perhaps $\sim O(10)$, times larger than
the orbital period.  However, when radiation pressure dominates gas
pressure, photon diffusion may cut in at relatively large scales,
particularly in regard to compressive modes (Agol \& Krolik 1998).
When that is the case, the multiplicative factor might be rather
smaller.  At the same time, not all of the energy has to be thermalized
by the turbulent cascade; other losses may occur.  Buoyancy, for
example, causes magnetic flux to rise to the surface of the disk and
escape.  Because the underlying force is gravitational, the
characteristic timescale for this motion is the orbital period, but
fluid drag and the necessity of draining some of the attached matter
can slow this process.  Thus, we write $t_{turb} = \tau_{turb} P_{orb}$
and expect that $\tau_{turb}$ will be between a few and $\sim 10$.

To determine the turbulence edge, we compare this cascade timescale to
the inflow time.   Well outside $r_{ms}$ in a thin disk, the inflow
time is always much longer than the turbulent cascade time.  Indeed,
this contrast is part of the separation of scales that makes it
reasonable to parameterize the stress by $\alpha$.  However, the inflow
time diminishes rapidly as the region of the marginally stable orbit is
approached from the outside.  One way to estimate the mean inflow time
in that regime is via the equation of angular momentum conservation as
applied to a statistically time-steady disk
\begin{equation}
\int d\phi \, \int \, dz T_{r\phi} = \dot M \Omega
       \left[1 - j_{in}/j\right],
\label{jcons}
\end{equation}
where $T_{r\phi}$ is the $r$--$\phi$ component of the stress tensor.
This equation may be rearranged to yield an estimate of the inflow
time:
\begin{equation}
t_{in} = {r^2 \int d\phi \, \int dz \, \rho \over
          \int d\phi \, \int dz \, \rho v_r r} =
          {r^2 \Omega \left[1 - j_{in}/j\right] \int d\phi \, \int dz \, \rho
         \over
         \int d\phi \, \int dz \, \langle B_r B_\phi/4\pi \rangle},
\label{inftime}
\end{equation}
where $\rho$ is the volume density, and we have assumed that all
the angular momentum transfer is due to magnetic stress.  Note that
this estimate is actually the inverse of the density-weighted mean infall
rate, rather than the density-weighted infall time.  It also assumes that
infall is limited solely by the rate at which angular momentum can be
removed.  It is most appropriate in the rotationally supported
part of the disk, and is wholly inapplicable inside $r_{ms}$ where,
by definition, rotation cannot prevent infall.

\begin{figure}

\plotone{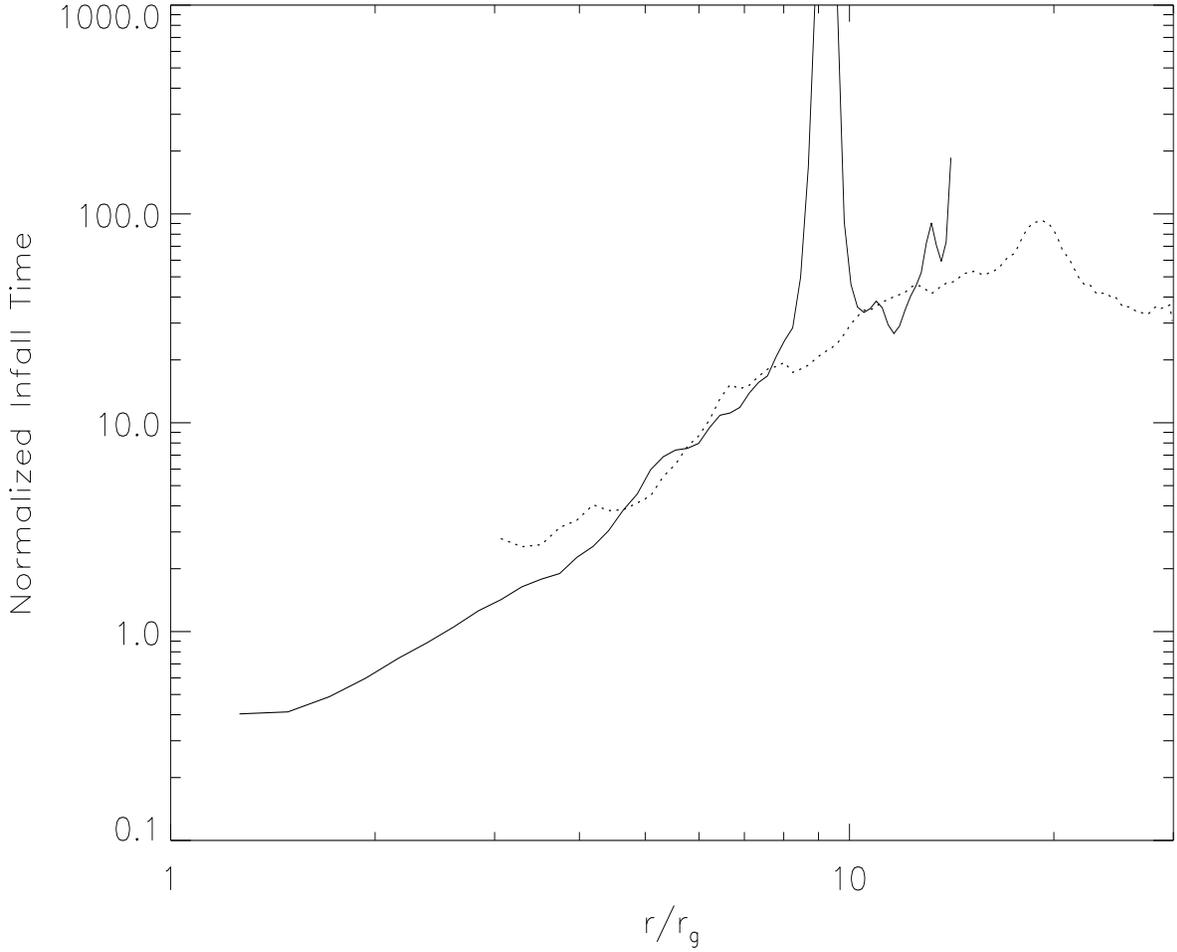}
\caption{Mean infall time in units of the orbital period estimated
two different ways.  The solid curve shows the density-weighted mean
value of $r/(v_r P_{orb})$ in a late-time snapshot from the high-resolution
initially-poloidal simulation of HK02; the dotted curve shows the
mean infall time as estimated by eq.~\ref{inftime} using data
from the same simulation and setting $j_{in} = 0.95j(r_{ms})$, as
indicated by the results of HK02.  The spike in the solid curve near
$r = 10 r_g$ and that curve's end near $r=15 r_g$ are both artifacts
of the finite mass in the simulated accretion disk; at large radii, its
material, in net, moves out.
\label{infalltime}}
\end{figure}

    In the conventional treatment (e.g. Abramowicz \& Kato 1989), the
stress is assumed to be $\simeq \alpha p$ for a constant $\alpha
\lesssim 1$ and pressure $p$.  Because the pressure decreases sharply
as the flow accelerates inward near the marginally stable orbit, the
stress falls equally sharply.  The factor in square brackets in the
numerator of the right-hand-side of equation~(\ref{inftime}), which
approaches zero near the inner edge, is then nearly cancelled by the
diminishing stress in the denominator.  However, when the stress is due
to MHD turbulence, it does not decline near $r_{ms}$, the ratio of stress
to pressure is not a constant, and the inflow time as estimated by
equation~(\ref{inftime}) becomes steadily shorter as the flow moves
toward the marginally stable region from the outside (see
fig.~\ref{infalltime}).  The inflow time as estimated by the torque
condition does not approach zero even at $r_{ms}$ because $j_{in} \neq
j(r_{ms})$ in the simulation.

Comparison of this estimate with one derived by computing a
density-weighted mean of $r/v_r$ shows that angular momentum removal
by magnetic stress is, indeed, the primary determinant of inflow in the
disk body, but infall no longer becomes angular momentum-limited
at radii inside $\simeq 4r_g$ (we define the gravitational radius
$r_g \equiv 2GM/c^2$).  Inside $r_{ms}$, continued inflow
can occur on the dynamical timescale without any change in angular
momentum; even in the region not far outside $r_{ms}$, little torque
is required, and other forces, even if small, can help push matter inward.
Primarily because of the approach of $j$ to $j_{in}$ , the ratio of
the infall time to the orbital period falls sharply inside $r=10r_g$,
dropping below 10 inside $r\simeq 7 r_g$.\footnote{
Even in statistically steady disks, local
fluctuations are so large that individual fluid elements can spend very
different lengths of time traversing a given radial span.  A more
complete treatment would therefore consider the probability
distribution for the infall time of matter at a given radius.} Because
the turbulent cascade time is perhaps $O(10)$ orbital periods, this line of
reasoning would suggest that the turbulence edge (in this simulation)
is near $\simeq 7 r_g$.

\subsection{Magnetic Field Structure}

   Another approach to determining the turbulence edge is to look for
a change in magnetic field structure that might result from the
turbulence-dominated to flux-freezing transition.   In the disk
body, where turbulence dominates, the magnetic field structure
is determined by a balance between stirring by the MRI and nonlinear processes
that transfer energy to smaller scales, where the energy is
ultimately dissipated.  Although
these nonlinear processes do not intrinsically lead to correlations
between different field components, the linear properties of the MRI
and the consistent sense of orbital shear impose a correlation between
$B_r$ and $B_\phi$ such that $\alpha_{mag} \equiv \langle B_r
B_\phi\rangle/\langle B^2 \rangle \simeq 0.25$ (Hawley, Gammie \&
Balbus 1996; HK01).  Indeed, it is these correlated magnetic
fluctuations that enable the MHD turbulence to serve as an angular
momentum transport mechanism.  On the other hand, where the flow is
dominated by infall, so that the field evolution is best described by
simple flux-freezing, orbital motion is the dominant factor and one
would expect the shear to create a larger $B_r$--$B_\phi$ correlation.
Thus, the radius inside which $\alpha_{mag}$ begins to grow above its
characteristic turbulent value can mark the place where transfer of energy
to small lengthscales becomes slower than infall.

\begin{figure}

\plotone{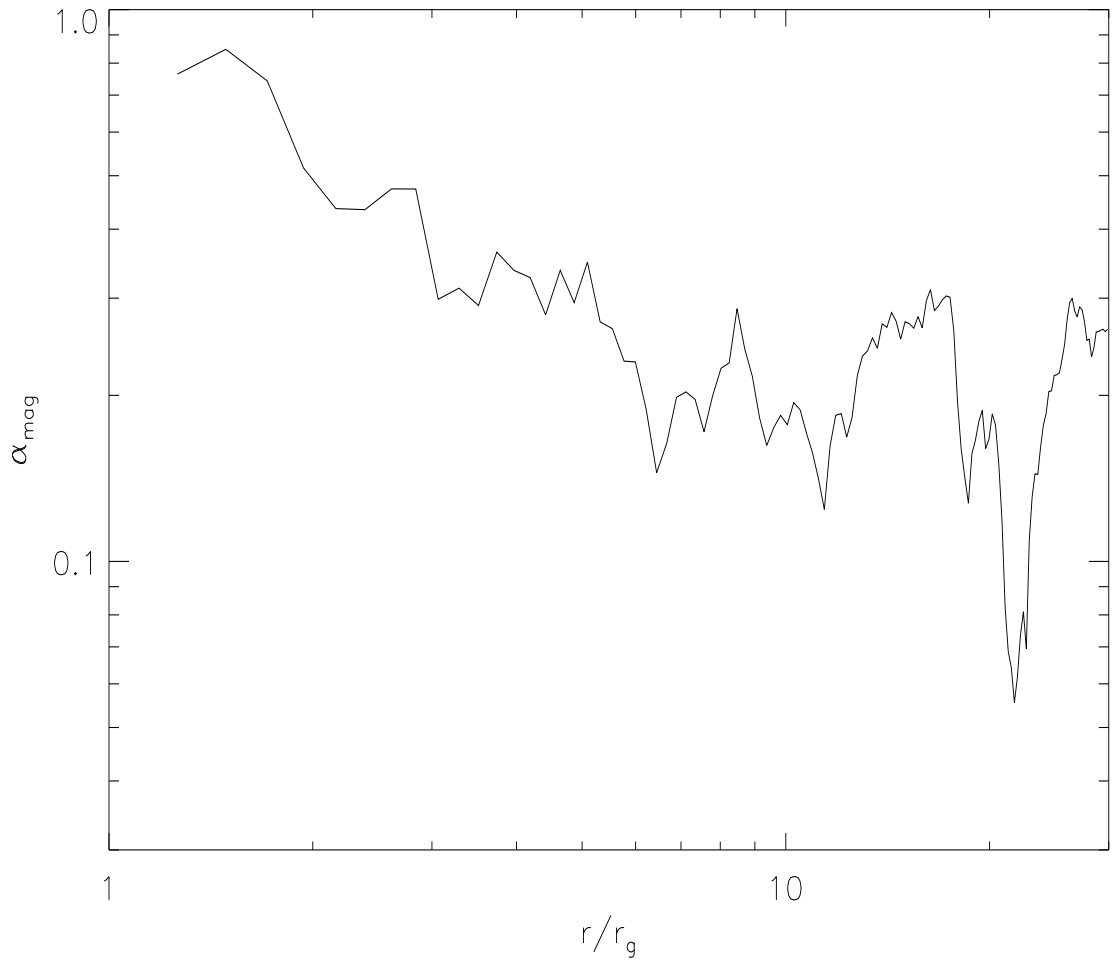}
\caption{The density-weighted ratio of magnetic stress to magnetic
energy density as a function of radius.  Data are taken from the same
set used for Figure~\ref{infalltime}.
\label{alphamag}}
\end{figure}

To examine this in detail in the initially-poloidal simulation of HK02,
we compute the density-weighted mean value of $\alpha_{mag}$ in
a single late-time snapshot as a function of radius (Fig.~\ref{alphamag}).
As the figure shows, $\alpha_{mag}$ begins to rise at radii less than
$\simeq 6r_g$.  The position of this rise fluctuates substantially
from time to time.  However, wherever the change begins, $\alpha_{mag}$
rises steadily inward until reaching a value near unity deep in the
plunging region.
In the initially-toroidal simulation of HK02, this transition occurs at rather
smaller radius, always close to 3$r_g$, but the rise inward of this point is,
if anything, sharper.

\subsection{Infall and turbulent velocities}

\begin{figure}

\plotone{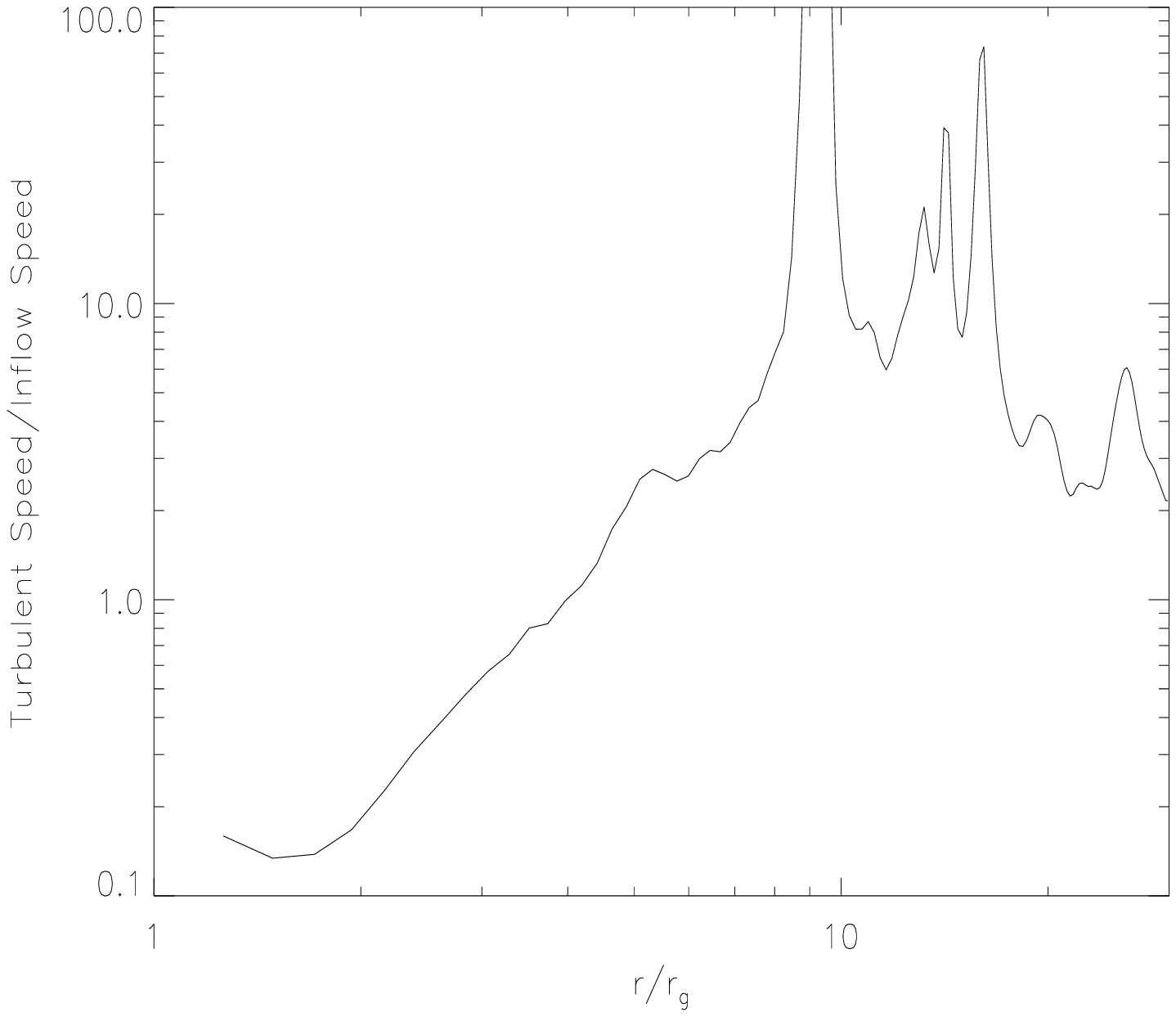}
\caption{The ratio $v_{r,rms}/\bar v_r$, where $\bar v_r$ is the
density-weighted mean radial velocity and $v_{r,rms}$ is the
density-weighted {\it rms} radial velocity fluctuation relative
to $\bar v_r$.  The data are from the same set used in
Figures~\ref{infalltime} and \ref{alphamag}.
\label{turbspeeds}}
\end{figure}

In the heart of the accretion disk, fully developed MHD turbulence
produces velocity fluctuations relative to the mean velocity that are
substantially larger than the mean accretion velocity itself.  As
the flow approaches the inner edge, however, the accretion velocity
increases relative to the fluctuations.  Thus, another measure of the
transition from the turbulence-dominated regime to the smooth infall
regime is the ratio of the random velocity to the mean.   As
Figure~\ref{turbspeeds} shows, this ratio is $\sim 5$ -- 10 in the
disk body, but begins diminishing near $r=8r_g$ and falls $\propto r^2$
all the way into the plunging region.  Near the inner edge of the
simulation, the turbulent fluctuations are only $\sim 0.1 \times$ the
mean radial velocity.

\subsection{Summary}

   Based on these estimates, we conclude that the turbulence edge in
initially-poloidal simulations is in the neighborhood of 6 -- $8 r_g$,
but may be closer to $r_{ms}$ when the field is initially toroidal.
For reasons discussed in HK02,
it is likely that the initially-poloidal case is closer to reality
than the initially-toroidal one, so one might expect the larger radius
to be more representative of real disks.

    If so, we may calibrate the factor $\tau_{turb}$ {\it post hoc}.
Assuming that the turbulence edge is at $r \simeq 7r_g$, we see
from Fig.~\ref{infalltime} that $t_{in} \simeq 7 P_{orb}$ at that
point.  By definition, $t_{in} \simeq t_{turb}$ at that radius, so we
conclude that $\tau_{turb} \simeq 7$, in line with our expectations.

     The position of the turbulence edge may be a
function of disk thickness $h$, and/or the accretion rate.  The
infall time $t_{in}$ is always $\sim P_{orb}$ in the inner disk, but it
increases to $\sim \alpha^{-1} (r/h)^2 P_{orb}$ in the disk body.
Therefore, the infall time must rise more rapidly near $r \simeq 10r_g$
in cold, thin disks than in the simulated disks, which have $h/r \sim
0.1$--0.2.  On this basis one would expect the turbulence edge to move to
smaller radius for smaller $h/r$.  Similarly, inward-directed pressure
gradients can be significant within a distance $h$ of $r_{ms}$, causing
the final plunge to begin at a larger radius for hotter disks.  However,
there is little room for the turbulence edge to move too much closer
to $r_{ms}$, so the dependence on $h/r$ might be relatively weak.

\section{The Stress Edge}

\subsection{General considerations}

   When matter crosses the event horizon of the black hole, it loses
all ability to communicate with the outside world, even with signals
traveling at $c$.  If the accretion flow were time-steady and
spherically symmetric, it would be easy to define an analogous surface,
{\it the stress edge},
where a more restricted loss of communication takes place: the surface
where dynamical communication ceases.  That is the surface on which the
inflow speed (as measured, for example, by a distant observer) exceeds
the magnetosonic speed (measured relative to the same observer).

   In real accretion flows, which are thoroughly non-steady and
non-symmetric, it becomes much harder to define such a surface.
Even in steady flows, asymmetry significantly complicates the issue.
Imagine, for example, a steady flow in which the magnetosonic surface
has a dimple.  It would be entirely possible for signals to travel
diagonally inward within the super-magnetosonic region, cross the
boundary into the sub-magnetosonic region, and travel outward from
there.  When that happens, the magnetosonic surface is no longer a
final boundary beyond which the flow loses causal contact with the
outside world.  The surface on which that takes place lies somewhere
closer to the black hole; its exact location depends on details of the
flow structure.

   If the flow is non-steady, which is the realistic case, individual
asymmetric structures like the one described in the previous paragraph
are the norm, but with the additional complication that they are
transient.  Causal paths need exist only for the time the signal
travels through them.  Time-variation might be so violent that
time-averages would erase any indication that such paths exist.  The
only way to determine these causal trapping surfaces for certain is to
trace all characteristics through spacetime to their end, either in the
black hole or back to the outside.  For all these reasons, the concept
of an average stress edge is perhaps the least well-determined of all the
edges discussed here.

    Giving up hope of defining this edge exactly, instead we attempt
to locate it approximately using two crude indicators: the ratio of
infall speed to magnetosonic speed, and gradients in the matter's
angular momentum and energy.

\subsection{Comparing the magnetosonic and infall speeds}

\begin{figure}

\plotone{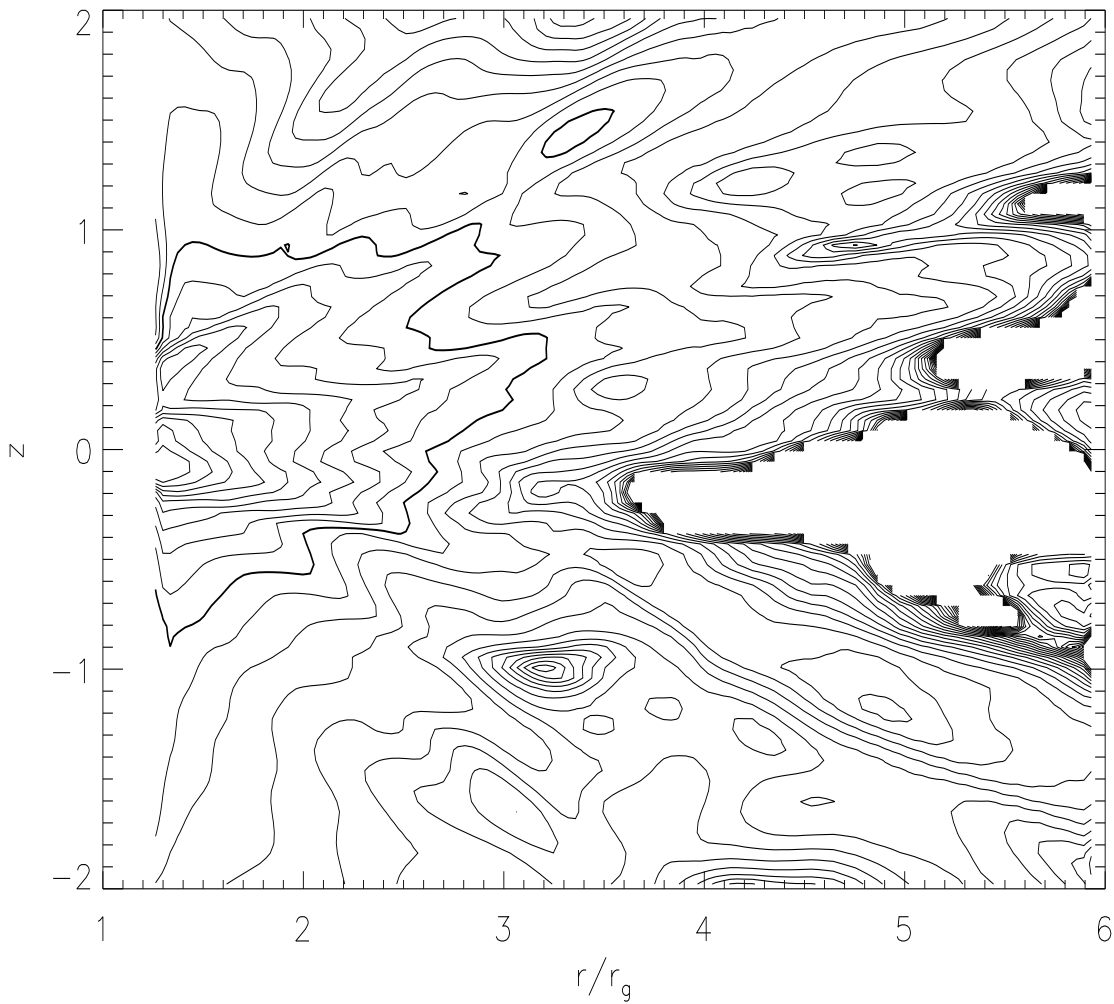}
\caption{Azimuthal average of the local ratio of infall speed to
magnetosonic speed at late time in the initially-poloidal simulation
of HK02.  Contours are logarithmic, spaced every 0.1 dex.  The
heavy contour cutting the equatorial plane near $r=2.5r_g$
marks the instantaneous azimuthally-averaged magnetosonic surface.
Where the contours disappear inside the disk body, the mean ratio of
infall speed to magnetosonic speed is less than $10^{-3}$.
\label{polvratio}}
\end{figure}

    As was just discussed, the point where the infall speed becomes
equal to the magnetosonic speed clearly demarks the stress
edge only in spherically-symmetric, time-steady flows.  Nonetheless,
we might look to this ratio as giving at least a crude estimate of
this edge's position.  Fig.~\ref{polvratio} shows an azimuthal
average of this ratio at a particular late-time instant in the
initially-poloidal simulation of HK02.  In this snapshot, the surface
within which the infall speed is
generally greater than the magnetosonic speed
falls roughly between 2.5 and $3r_g$ between the
midplane and $z\simeq +r_g$, but slants radially inward below
the equatorial plane.  Thus, from this sort of data, one might
imagine the stress edge as occurring at a variety of radii between
2 and $3r_g$, depending on altitude.  The asymmetry between regions
above and below the midplane is, of course, time-variable.  Inside
the disk body (i.e., $r \geq 3.5r_g$ and $|z| \lesssim 0.1r$), the
mean infall speed is a tiny fraction of the magnetosonic speed.

Next, we consider this ratio in the equatorial plane as a function of
time.  After $t=1000$, when the accretion flow has established a
quasi-steady state, the location where $v_r$ becomes super-magnetosonic
remains inside $r_{ms}$, but varies rapidly, from a minimum at
$r=2.3r_g$ out to a maximum near $r=2.9r_g$.  We expect that there are
also similarly large fluctuations in the location of the magnetosonic
surface away from the equatorial plane.

\subsection{Specific angular momentum}

\begin{figure}

\plotone{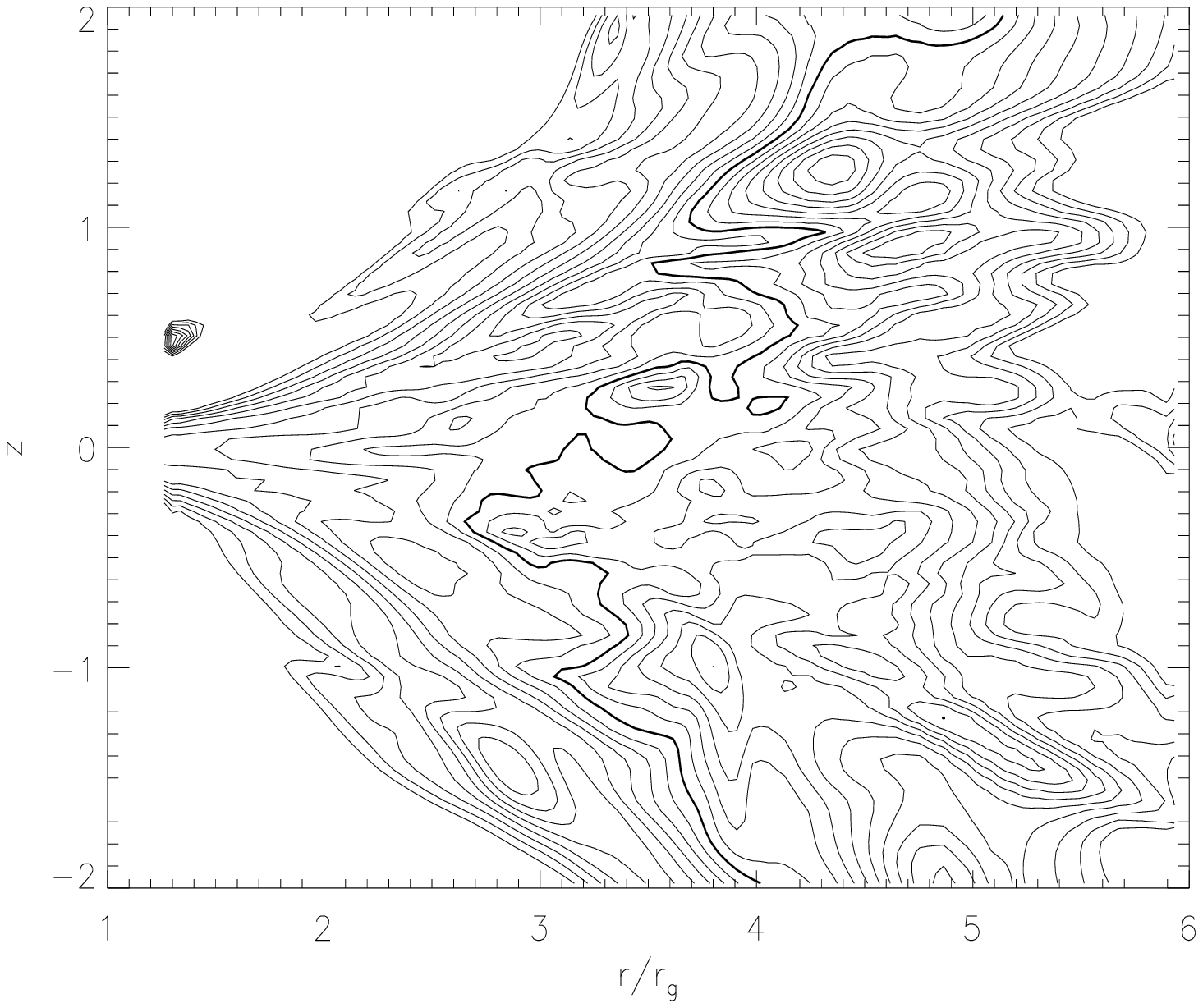}
\caption{Azimuthal average of the the specific angular momentum
at the same late time as Fig.~\ref{polvratio} in the initially-poloidal
simulation of HK02.  Contours are linear, spaced by 0.03 in units
of $\sqrt{2}GM/c$.  The heavy contour marks the specific angular momentum
(2.60) of the marginally stable circular orbit.
\label{polangmom}}
\end{figure}

    A contrasting perspective is given by the variation in
specific angular momentum across the accretion flow (fig.~\ref{polangmom}).
As this figure shows, $\langle \partial j/\partial r\rangle > 0$
all the way to the inner edge of the simulation problem area at
$r = 1.25r_g$.  In the equatorial plane, the specific angular momentum
continues to fall at least as far inward as $r\simeq 1.5r_g$.
Away from the midplane, the azimuthally-averaged
poloidal velocity field angles toward the equatorial plane, but
not as steeply as the angular momentum contours; hence, the
angular momentum of individual fluid elements falls as they
flow inward.  If this figure were our only diagnostic
of the stress edge, we might conclude that this edge lies
inside the smallest radius treated in this simulation, i.e.
at less than $1.25r_g$.

\subsection{Relation to turbulence edge}

     For a third estimate of the stress edge's location, we examine its
relation to the turbulence edge.  The stress edge must be inside the
turbulence edge for it must occur where the net inflow velocity dominates
over the velocity fluctuations.   Moreover, because conditions at the turbulence
edge provide a sort of boundary condition for the relatively predictable
behavior in the flux-freezing region, we can use a simple analytic argument
to estimate how far inside the turbulence edge the stress edge occurs.  In
essence, this argument is the same as the one employed in Krolik (1999a).

\begin{figure}

\plotone{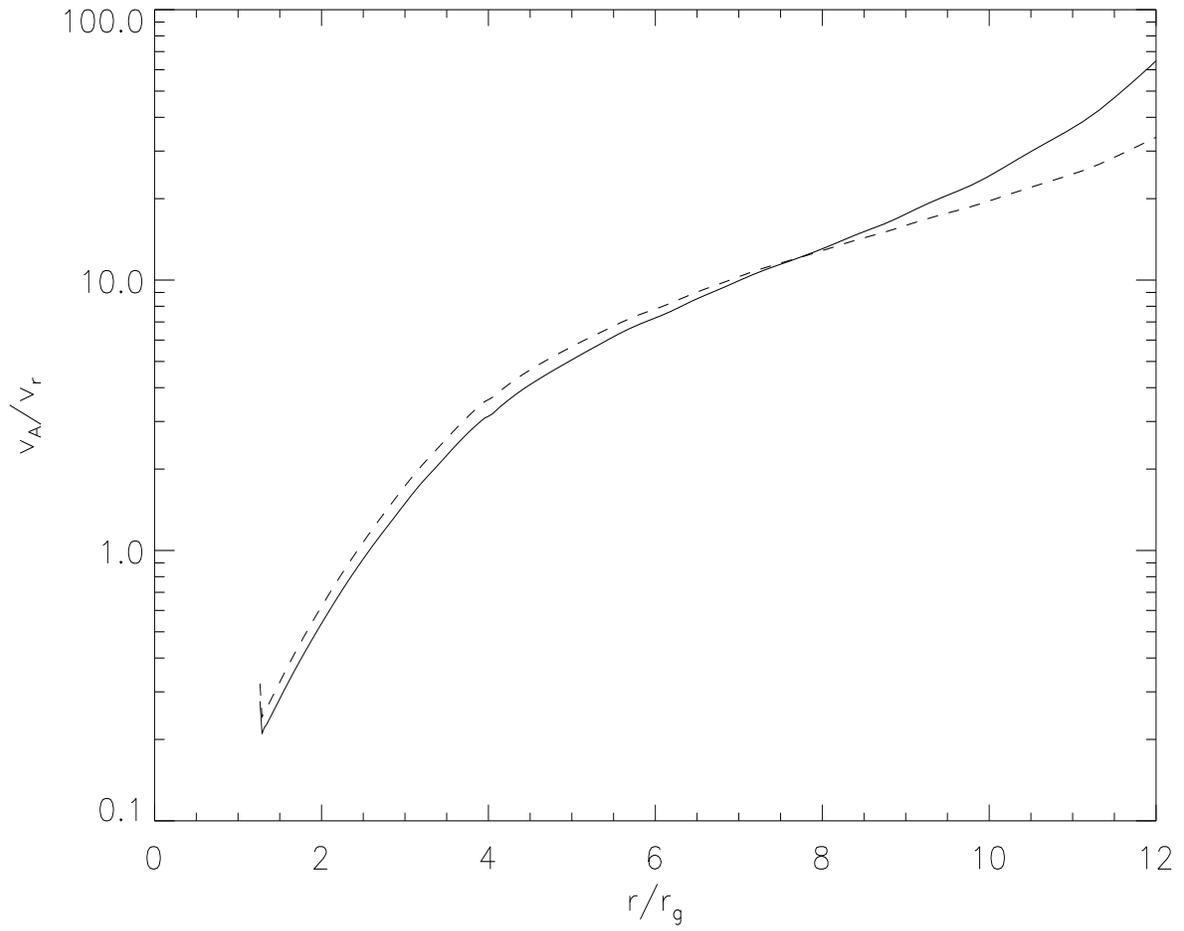}
\caption{Plot of the ratio of the Alfv\'en speed to the infall
velocity (solid line), and the right hand side of equation~(\ref{velratio})
(dashed line).
\label{valfvr}}
\end{figure}

Our reasoning begins with equation~(\ref{jcons}), evaluated
at the turbulence edge and assuming that all the stress is magnetic.
This condition fixes the magnetic field intensity at the turbulence
edge.  Extrapolating the field strength to smaller radii with the
assumption of flux-freezing then gives an estimate of the Alfv\'en
speed.  Writing all quantities evaluated at the turbulence edge with a
subscript $*$ and introducing mass-conservation in the form
$2\pi rh \rho v_r = \dot M$, we find
\begin{equation}\label{velratio}
{v_A \over v_r} = \left\{ {2 \over \alpha_{mag*}} 
{\langle B^2 \rangle \over \langle B^2 \rangle_*}
{rh \over r_* h_*} {v_{\phi*} \over v_r} 
\left[1 - j_{in}/j_*\right]\right\}^{1/2} ,
\end{equation}
where $v_\phi$ is the azimuthal speed.  To verify equation~(\ref{velratio}), we
plot (in fig.~\ref{valfvr}), the left hand and right hand sides of this
equation for the inner region of the initially poloidal simulation from
HK02.  The values are averaged over height, angle, and time from
$t=1000$ to 1830 (the endtime).   If we set $\alpha_{mag*} = 0.3$,
we find that the simulation data give a very close match to
this analytic prediction for $r_{*}$ between $\sim 6$ and $8r_g$;  The
figure uses $r_*=6.3r_g$, a number well within the range previously
estimated for the turbulence edge.

Supported by this check, we may then use equation~(\ref{velratio}) to estimate
the position of the magnetosonic surface based on conditions at the
turbulence edge, once again using that surface as a rough guide to
the location of the stress edge.  Because the
simulations also suggest that the height-integrated magnetic energy
doesn't change much as a function of radius in this region, the
definition of the stress edge $r_s$ based on equation~(\ref{velratio})
simplifies to
\begin{equation}\label{simpr_s}
r_s \simeq \left[ \alpha_{mag*}{j_{in}/j_* \over 1 - j_{in}/j_*}
{v_r (r_s) \over v_\phi (r_s)}\right]^{1/2} \simeq 
\left[{v_r (r_s) \over v_\phi (r_s)}\right]^{1/2}.
\end{equation}
The simplified form reflects the simulational results that $\alpha_{mag*}
\simeq 0.3$ and $j_{in}/j_* \simeq 0.8$.  This form demonstrates that
$r_s < r_{ms}$ because $v_r/v_\phi$ is generally still a small number
until well inside $r_{ms}$.  For example, for the data shown in Figure~1 the
mass-weighted mean $v_r/v_\phi \sim 0.1$ at $r_{ms}$,
so that $r_s \simeq r_*/3 \simeq (2$ -- $2.5)r_g$.
We emphasize, however, that the location of the turbulence edge
fluctuates with time, causing movement of the stress edge.  A relation
such as equation~(\ref{velratio}) holds in an approximate and
time-averaged sense.
 
\section{The Reflection Edge}

The edges discussed above are dynamical.  Other interesting edges
have to do with the innermost places from which observable photons
may emerge.

    Many accreting black holes exhibit Compton reflection features.
These are broad bumps in the spectrum from $\simeq 10$ to $\simeq 50$~keV.
All photons can be reflected by Compton scattering from ionized material,
but those with energy below $\simeq 10$~keV are more likely to be
absorbed by a variety of photoelectric processes, while those with energy
$\sim 100$~keV or more lose energy to Compton recoil in the very process of
reflection (Lightman \& White 1988).  To produce a Compton
reflection feature, accretion disks must satisfy two criteria: 
they must be optically
thick to Compton scattering, and their matter must not be too ionized.

    Two very similar criteria regulate Fe K$\alpha$ production in
disks.  This emission line can be generated when hard X-rays illuminate
matter containing unstripped Fe atoms and ions.  When the Fe is less
ionized than H-like or He-like, the line is produced by fluorescence;
in the higher ionization stages, the process is actually recombination,
but it scales in identical fashion with illuminating flux until most
Fe atoms are fully stripped of their electrons.

   Because the Fe K$\alpha$ line and the Fe K-edge that dominates the
low-energy end of the Compton reflection feature can be shifted in energy
by general relativistic effects, their location in X-ray spectra
can provide direct diagnostics of orbital properties deep in the relativistic
potential.  It is therefore very important to correctly identify their origin.

   Although it is common
to assume that this {\it reflection edge} is identical to the marginally
stable orbit, Reynolds \& Begelman (1997) pointed out that this is
not necessarily the case because material farther in could be
illuminated by hard X-rays generated elsewhere.  On the other hand,
Young et al. (1998) questioned whether there would be enough unstripped
Fe in the lower-density material in the plunging region to efficiently
produce fluorescence photons.  The fractional abundance of unstripped
Fe is controlled by the ratio $J_x/\rho$,
where $J_x$ is the mean intensity of X-rays with energy $> 7$~keV.
When $J_x/\rho$ is too great, the efficiency of K$\alpha$ production
falls because an increase in the illuminating flux no longer produces
any increase in line photon emission.  In addition, because the Fe
K-edge photoionization opacity of matter with a solar abundance of Fe
is approximately the same as the Thomson opacity, illuminating
photons are only fully used when the column density is great enough
to make the matter Compton thick.  Thus, the falling density in
the plunging region reduces the K$\alpha$ emission efficiency for
both these reasons.

    One of the most noteworthy results
of the MHD simulations is to show that, particularly in the plunging
region, accreting matter fluctuates widely in density, in both space
and time.  Regions separated by less than the azimuthally-averaged
scale height can differ in density by factors of 10--30.  As a result,
the optical depth at fixed radius can vary in azimuth by factors of
2--3.  A treatment that assumes all the accreting
matter has the same
mean density and optical depth could therefore miss important effects.

    In our simulations, we have no way of estimating the hard X-ray
intensity, and, because we have no definite unit of density, we can
measure the gas density and column density only in relative terms.
However, by normalizing the mass inflow rate to the Eddington rate,
we can make a correspondence between the density in the simulation and
a physical density.  In physical units,
$\dot M = 1.7\times 10^{17}(\dot m /\eta) M/M_\odot {\rm gm\ s^{-1}},$
where $\dot m$ is the mass inflow rate in Eddington
units and $\eta$ is the ``efficiency'' factor, i.e.,
the fraction of the rest-mass energy available for radiation.

\begin{figure}

\plotone{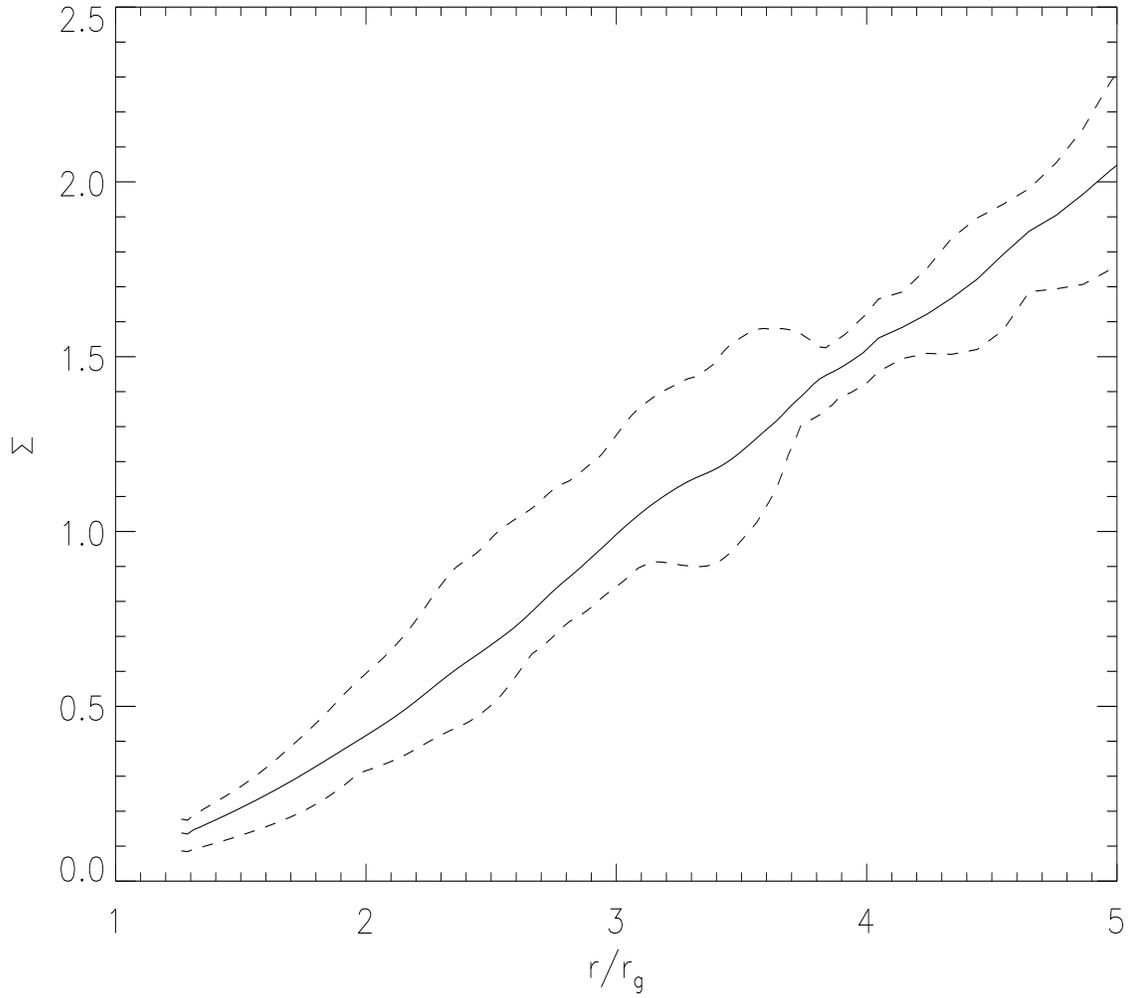}
\caption{Plot of surface density $\Sigma$ as a function of radius
in inner disk.  The solid line is the azimuthally-averaged value,
normalized to unity at $r_{ms}$.  The dashed lines are the maximum and
minimum values at each radius.
\label{sigma}}
\end{figure}

Figure \ref{sigma} shows the value of the surface density $\Sigma$ as
a function of radius for the inner region of the poloidal field
simulation of HK02.  The solid line is the azimuthally-averaged value
and the dashed lines are the minimum and maximum values.  The inflow
stream has a prominent $m=1$ spiral pattern, and the minima and
maxima provide a measure of the variation with angle.  The values are
normalized to the average value at $r_{ms}$.

The surface density $\Sigma$, and with it the optical depth, decline
smoothly inward, dropping by a factor of $\sim 10$ from $r_{ms}$ to the
inner boundary and increasing roughly proportional to $r^{3/2}$ out to
$r=10r_g$.  The drop in the plunging region is less severe than in
Figure 1 of Reynolds \& Begelman (1997) because they make the
artificial assumption that the infall velocity goes to zero at
$r_{ms}$.  In contrast, we observe no sharp change because our infall
velocity varies smoothly across $r_{ms}$.

The initially toroidal simulation from HK02  shows a steeper inward
decline in $\Sigma$, which is roughly proportional to $r^{3}$ out to
$r=10$.  Even here, however, there is no sudden break or change in
character at $r=r_{ms}$.

   Employing the conversion factor defined above, we find that the average
$\Sigma$ at $r_{ms}$ in the poloidal simulation is
$14 (\dot m /\eta)$~gm~cm${}^{-2}$.  The corresponding mean Thomson optical
depth is $\tau_T \simeq 5 (\dot m/\eta)$.  In other words, at high
accretion rate, all of the stably-orbiting part of the disk and some of
the plunging region are Thomson thick, but at low accretion rate, even
the inner part of the stable disk may be optically thin.  The
reflection edge may therefore lie on either side of $r_{ms}$, depending
on the accretion rate.  This conclusion
is at variance with standard analytic models for accretion disk equilibria
because, unlike them, we allow for the smooth acceleration in the radial
inflow speed that begins outside the marginally stable orbit.

   Although the mean column density does decrease inward, the maximum
volume density in the inflow spiral remains
roughly constant until $\sim 2r_g$, before dropping.  Of course, the
maximum fills an increasingly small volume, but assuming a constant
incident X-ray flux, the persistence of local high-density regions
means that there can be some material deep in the plunging region
that is no more highly-ionized than it was at the marginally stable orbit.

Finally, we comment that the reflection edge is, in some respects, the
nearest analog to what one might ordinarily define as the disk's inner
edge.  It is the place where the integrated surface density has declined
enough that the disk becomes effectively transparent.

\section{The Radiation Edge}

    The previous section discussed photon reflection; no reflection can
happen, of course, unless photons are generated to be reflected.  Thus,
we are led to the last edge we consider, the {\it radiation edge}, the
innermost radius from which significant radiation emerges.  There are two main
considerations in establishing this edge:  the heating that takes place
in the plasma, and the rate at which that heat can be radiated away.

\subsection{The dissipation distribution}

    Outside the stress edge in a time-steady disk, the amount of
energy available for local
dissipation is the difference between the net rate of change in
the matter's binding energy and the energy that it loses by doing
work through stresses on adjacent matter (eqn.~1).  Conventional
Novikov-Thorne models assume a boundary condition on the stress that
forces dissipative heating to end at the marginally stable
orbit (Novikov \& Thorne 1973, Page \& Thorne 1974).  As first suggested
by Krolik (1999a) and confirmed in the simulations of HK01 and HK02, stress
due to MHD turbulence continues right through the marginally stable
region and well into the plunging region.  Consequently, it cannot be
assumed that dissipation---and therefore radiation---end abruptly at
$r_{ms}$.

   One way that the dissipation distribution may be altered from the
Novikov-Thorne prediction is for the extra stresses in the inner part
of the accretion flow to convey energy coherently from plunging matter
back to the disk proper, where it can then be dissipated.  Agol \& Krolik
(2000) showed how a stress expressed at $r_{ms}$ translates into
an extended dissipation distribution at more distant radii.  Close
outside $r_{ms}$ the dissipation per unit area drops very rapidly with
increasing radius; at somewhat larger radii, it scales $\propto r^{-7/2}$.

   The Agol \& Krolik solution refers to the vertically-integrated
dissipation rate.  Because it is based only on energy conservation,
it has nothing to say about the vertical distribution of that dissipation
within the disk.  For the purpose of making more specific spectral
predictions, that distinction is important.  In particular, the
radial emissivity of reflection features (as discussed in the previous
section) depends on the coronal dissipation rate.  We will return to
this point in \S 6.

    A second alteration to the classical dissipation distribution
may occur in the plunging region itself.  Independent of how much
energy may be transferred from there to the disk body, there can be
very large local energy exchange between adjacent fluid elements.  For
example, in the initially-toroidal simulation of HK02,
there are places in the plunging region where material with binding
energy twice the mean finds itself a small fraction of a radius from
material that has very nearly zero binding energy.  Some of the
energy involved may find itself dissipated.  In addition, stripes of
oppositely-directed magnetic field can be brought very close together,
offering opportunities for rapid reconnection.  Because the field
energy is large compared to the pressure in this region, the heating
associated with reconnection could be quite important.

\subsection{Photon generation and escape}

     The stress distribution in the inner region of an accretion
disk also cannot automatically be transformed into a radiation distribution.
Any of several effects could
cause inefficiency in the transformation of energy into heat and thus
into escaping photons.  One possibility, which has been the focus of a
great deal of attention in recent years, is that the primary
dissipation mechanism heats the ions, but not the electrons (Rees et
al. 1982; Narayan \& Yi 1995).  In the present discussion, we will not
consider that possibility, except to note that to answer the question
of which species receives the majority of the heat will require
detailed consideration of the small-scale processes that
terminate the nonlinear cascade of the MHD turbulence.  Here we will
assume that $T_{ion} = T_e$.   Another plausible alternative channel
for energy loss is mechanical work.  This channel has two branches:
organized outflows such as winds and jets, and simple $p \, dV$ work
within the accretion flow.  A third possibility is that the inflow may
be so rapid that there is not enough time for the matter to generate
photons, and for the photons to diffuse out, before the fluid reaches
the black hole (Abramowicz et al. 1988).  Here we will concentrate
on the latter three issues:
adiabatic cooling, photon generation, and photon diffusion.

     One might expect adiabatic cooling to  become more important than
dissipative heating when the inflow time becomes shorter than the time
to dissipate the MHD turbulence, i.e., inside the turbulence edge, but
there are two loopholes in this line of reasoning.  The first
is whether the gas actually falls in density as it plunges inward.  If
all that happened were an increase inward of $|v_r|$, a sharp density
drop would be inevitable.  However, as seen in the simulations of HK01
and HK02, and as discussed above in \S4, the density in the spiral
inflow can remain higher than what might be predicted.  Convergence to
the equator and into a nonaxisymmetric spiral can compensate for
inward acceleration.

    The second loophole is that the turbulence edge is defined
specifically in terms of the turbulent cascade of magnetic energy to
short lengthscales.  Other, more rapid, forms of dissipation can also
occur, such as shocks and magnetic reconnection.  In particular, in the
simulations of HK02 there are enough weak shocks that the specific
entropy (defined as $p/\rho^{5/3}$) rises by roughly a factor of 5 from
$r=6r_g $ to $r=1.25r_g $ in the equatorial plane.  As a result, the
temperature actually rises slightly with decreasing radius in this
region.  Because this simulation does not capture several other
sources of heating (notably numerical magnetic reconnection), the
heating recorded is probably an underestimate.

     In most---but not all---circumstances, the photon generation time
is shorter than the photon diffusion time.  To illustrate the range of
possibilities, we consider two cases, one in which the primary
radiation mechanism is bremsstrahlung, the other in which the
primary mechanism is inverse Compton scattering.

     First consider bremsstrahlung.  It is slowest relative to the infall
time if the gas is essentially in free-fall.  When that is a good description,
\begin{equation}\label{bremstime}
{t_{brems} \over P_{orb}} = 
   {1 \over 9 \alpha_{fs} [\ln(kT/m_e c^2) + 1.5]} {\eta \over L/L_E}
   {h \over r} {v_r \over v_{ff}},
\end{equation}
where $t_{brems}$ is the characteristic time to radiate the gas's thermal
energy by bremsstrahlung, the numerical factors incorporate an approximation
to the relativistic bremsstrahlung Gaunt factor, $\alpha_{fs}$ is the
fine-structure constant, and $v_{ff}$ is the radial free-fall speed (see, e.g.,
Krolik 1999b for the bremsstrahlung radiation coefficient written in this
notation; note, too, that this expression assumes a pure electron-proton
plasma).  Thus, when the accretion flow is geometrically thick and close to
free-fall, the bremsstrahlung photon generation time is longer than
an infall time in the plunging region, particularly for low accretion
rate relative to Eddington.  However, when the flow is slower, the photon
generation time can easily be shorter than $t_{in}$.

    In the initially-poloidal simulation of HK02, $h/r \simeq 0.1$ throughout
the plunging region.  The time- and mass-averaged mean inflow velocity $v_r$
varied from $0.01v_{ff}$ at $r =7r_g$ to $0.35v_{ff}$ at $r=2r_g$.  At
late time, the mass-averaged temperature ($kT\propto p/\rho$) rises
inward from $0.001c^2$ at $r=10r_g$ to $0.004c^2$ at $r=2r_g$.  Using
equation~(\ref{bremstime}), we would then predict that the time required to
radiate the gas's thermal content by bremsstrahlung is
$0.008 \eta (L/L_E)^{-1} P_{orb}$.  Thus, bremsstrahlung cooling
is rapid compared to the dynamical time so long as the accretion rate is
greater than $\sim 10^{-2}\times$ the Eddington rate.

    Bremsstrahlung is not the only plausible cooling mechanism;
Compton cooling generally dominates when the electron temperature is
substantially greater than the Compton temperature, 1/4 the
intensity-weighted photon energy, and the electron density is relatively
low.  Compton cooling is especially likely to be important when dissipation
occurs within the plunging region.  The characteristic energy loss time for
this mechanism is
\begin{equation}
{t_{Compt} \over P_{orb}} = {9 \over 32\pi}{m_e \over \mu_e}
        {(r/r_g)^{1/2} \over L(x)/L_E},
\end{equation}
where $\mu_e$ is the mass per electron (generally $\sim m_p$)
and $L(x)/L_E$ is the luminosity produced inside radius $x$, relative
to Eddington.  Unless $L/L_E \ll 1$, the Compton time remains much
shorter than an orbital period throughout the (inner) portion of the disk
that radiates nearly all the light.

   On the other hand, depending on location, the diffusion time can be
either much longer than an orbital period or rather shorter.  In the
disk body, the diffusion time is generally $\sim \alpha^{-1} P_{orb}$.
However, as we have already estimated, the surface density, and hence
vertical optical depth of the disk, diminishes smoothly through the
inner part of the accretion flow, and, depending on parameters, the flow
may become optically thin to Compton scattering either inside or outside
$r_{ms}$.  Where the disk is optically thin, photon scattering is unimportant;
where it is optically thick, the diffusion time is
$\sim \tau_{T}^2 (h/r) (r/r_g)^{-1/2} P_{orb}$.

    On the basis of these estimates, we conclude that the radiation
edge of sub-Eddington disks is primarily determined by the total (i.e.
turbulent plus shock plus reconnection) dissipation distribution.
Where inflow is slow enough that the optical depth is large, $t_{in}$
is longer than the radiation time; where inflow is faster, the optical
depth falls sufficiently to permit efficient photon escape.  Thus, if
the flow inside the turbulence edge is laminar, the radiation edge will
be located at most a short distance inside the point where most dissipation
ceases.   If there is dissipation at small radii, the disk could remain
bright close to the black hole.

\section{An observational application: Fe K$\alpha$ profiles}

    The emissivity of reflection features---both the ``Compton bump"
and Fe fluorescence---is determined by a combination of the different
mechanisms that set the locations of the radiation and the reflection edges.
Most efforts to infer the reflection emissivity have concentrated on
Fe K$\alpha$, so we will focus on that feature here.

    Almost all phenomenological inferences of the K$\alpha$ emissivity
have assumed that the inner edge of emission is identical with $r_{ms}$.
For example, Nandra et al. (1998), fit their ASCA data from a
number of Seyfert galaxies to a model for which the emissivity $j_{K\alpha}
\propto r^-\beta$ for $r \geq 6GM/c^2$ ($r_{ms}$ for $a/M=0$)
and zero inside that radius.  For their sample, they found a mean value of
$\beta \simeq 2.5$.  Wilms et al. (2002) fit their K$\alpha$ profile from
MCG --6-30-15 to a similar model, but with the inner radius and $a/M$ variable.
They found that $4.3 \lesssim \beta \lesssim 5$, and,
by arguing that the inner edge of emission could be identified with $r_{ms}$,
placed a lower bound on $a/M$.

    Now that we can begin to distinguish between the radiation and reflection
edges and the marginally stable radius, it is necessary to approach these
inferences more circumspectly.  As has already been discussed in \S 5,
the coronal radiation edge---the version
of the radiation edge relevant to driving X-ray fluorescence---could fall
on either side of $r_{ms}$.

    Whether reflection ceases inside or outside $r_{ms}$ depends largely
on the accretion rate.  Thus, one could interpret extremely broad and
red-shifted Fe K$\alpha$ profiles in either of two generic ways:

\noindent {\it High accretion rate:} If $\dot m/\eta \sim 1$, the
reflection edge can be well inside $r_{ms}$.  In this case, the lower
bound on $a/M$ suggested by Wilms et al. (2002) would not apply because
many of the observed line photons could be created at $r < r_{ms}$.
Reflection requires irradiation, and to have sufficiently strong irradiation
in the plunging region, there must be either violent reconnection in that
zone or strong coronal emission
just outside $r_{ms}$.  As shown by Agol \& Krolik (2000), when extra
torque is applied to the disk at some radius $r_t \geq r_{ms}$, the
associated supplemental dissipation is concentrated very tightly at
radii just outside $r_t$.  In terms of Boyer-Lindquist coordinates,
the dissipation rate falls especially steeply outside $r_t$ when $a/M$
approaches unity.

\noindent {\it Low accretion rate:} If $\dot m/\eta \ll 1$, the
reflection edge is likely to move well outside $r_{ms}$.  In the
picture of accretion disks commonly adopted for line profile-modeling,
the column density of matter rises so sharply near $r_{ms}$ that disks
are all optically thick at
any radius $r \geq r_{ms}$.  However, as Figure~\ref{sigma} shows,
this is not so; the rise of optical depth with radius is relatively
smooth, so that disks can easily be optically thin well outside
$r_{ms}$, especially when $\dot m/\eta$ is small.  In this case,
the lower bound on $a/M$ posed by the prerequisites for creating this
sort of profile become quite stringent, and might become impossible to
satisfy.

    Both interpretations share one conclusion: in order to excite
enough K$\alpha$ emission at small enough radius to replicate the observed
broad profiles, it is necessary to extend the range of disk (coronal)
radiation well inward from the point where the Novikov-Thorne model
predicts that dissipation becomes weak.  If the K$\alpha$ photons were excited
by hard photons produced locally whose emissivity followed the Novikov-Thorne
dissipation distribution, $j_{K\alpha}$ would be very small near $r_{ms}$,
rise to a peak at significantly larger radius---in terms of Boyer-Lindquist
radii, $r_{peak} \simeq 1.6 r_{ms}$ for $a/M = 0$, and $1.2r_{ms}$ for
$a/M = 0.998$---roll over gradually, and then ultimately fall $\propto r^{-3}$.
The sorts of crude fits already in hand make it clear that this sort of
pattern for K$\alpha$ emissivity will not do---there must be additional
emissivity at small radii.

\section{Conclusions}

    The arguments presented here have shown how the simple term ``disk
inner edge" fragments into multiple meanings when accretion flows around
black holes are examined carefully and quantitatively.  At least four
different definitions might be interesting, two referring to dynamical
properties and two to observational ones.  Moreover, even when one
searches for an inner edge that is clearly defined conceptually, in a
real disk what one is likely to find is a fuzzy, asymmetric border that
varies in time.

    The outermost inner edge is likely to be the turbulence edge, the
place where the magnetic field ceases to be described by a balanced
turbulent cascade and is better thought of as evolving through flux-freezing
as the plasma that carries it flows inward.  If the initially-poloidal
simulation of HK02 is any guide, this occurs at 6--$8r_g$ from the
center of the black hole.  This edge marks an important transition in
the character of disk dynamics, from the familiar regime in which
$t_{in} \gg t_{th} \gg P_{orb}$ ($t_{th} \sim \alpha^{-1}P_{orb}$ is
the thermal time) to a quite different one in which all three
timescales are comparable.

    The other inner edge relevant to dynamics is the stress edge:
matter inside this edge cannot communicate energy or angular momentum
to matter outside.  We demonstrated that the stress edge should,
in general, lie well inside the turbulence edge, but it is difficult
to locate unambiguously, as it is particularly strongly
affected by time-variability and departures from symmetry.  The
position of this edge is central to accretion studies because it
regulates what are arguably the two most important parameters of the
system: the total energy and angular momentum removed from the
accreted material before it enters the central black hole.  Note that
because the stress edge is well inside the turbulence edge, much of
the energy released by accretion occurs in circumstances quite
different from those in the disk body.

    The reflection edge is the innermost material capable of producing
a significant Compton reflection feature or Fe K$\alpha$ fluorescence.
Because of the large fluctuations in gas volume and column density that
occur in the plunging region, this edge may not even be a clean surface;
rather, it may be that reflecting material breaks up into ``islands"
which become smaller and less common toward smaller radii.  Normalizing
the column density of matter in the simulations of HK02 in terms of
the accretion rate in Eddington units, we find that, on average, the
radius where the matter becomes thin to Thomson scattering could
lie either inside or outside $r_{ms}$; higher accretion rates lead
to reflection edges at smaller radii.

    The last of the four inner edges is the radiation
edge, the smallest radius from which significant luminosity
originates.  This is difficult to locate on the basis of current
simulations because they do not include either explicit dissipation
or radiation.  This edge is the most dependent on the detailed energetics
of the accretion flow, both with respect to dissipation and 
photon escape.  Estimates based on current work indicate that,
like the reflection edge, it could lie on either side of $r_{ms}$,
depending on circumstances.  It is entirely possible that
the disk can remain bright quite close to the black hole.

    Although the simulations done so far are not ideally suited to 
determination of the inner edges relevant to observations,
the dynamical
features of the simulations should be more robust.  One concern with
the present simulations is that the accretion flow originates in an
initial torus centered only $10 r_g$ from the black hole.  Although the
dynamical disk edges occur well within the inflow region of the
subsequent evolution, it is possible that specific details are
influenced by the finite torus size.  

The simulations done to date have also been restricted to relatively
thick flows, with $h/r \approx 0.1$.  The specific locations of the
various edges depend only weakly on disk thickness.  The turbulence edge
likely moves inward slightly for smaller $h/r$.  The location of
the stress edge probably does not depend directly on $h/r$.
Similarly, because surface density is independent of $h$, the location
of the reflection edge depends more on factors such as the net accretion rate.
Decreasing $h$ tends to move the radiation edge inward, we
expect, because, for fixed surface density, the diffusion time is
$\propto h$.  Future simulations should be able to refine these estimates.

In any case, while it is clear that any definition of a disk inner edge
will lie in the vicinity of the marginally stable orbit, there is
no reason why any of these inner edges should coincide {\it precisely}
with $r_{ms}$.  Depending on which physical concept is under consideration,
any particular inner edge might be a factor of 2--3 inside or outside
$r_{ms}$---and its position relative to $r_{ms}$ can easily change
as a function of time.

   These facts have important implications with regard to the use of
disk models as the basis for interpreting observations as
probes of accretion in strong gravity.  In virtually all previous
efforts to interpret Fe K$\alpha$ profiles, the line shape was fit to a model
that assumed the emissivity was a power-law in radius cut off abruptly
at $r_{ms}$.  However, as Reynolds \& Begelman (1997) first pointed
out, the reflection edge is not necessarily tied to the marginally
stable orbit; as we have argued, it could as easily lie outside as
inside that radius.  Similarly, fits to the thermal portion of black
hole accretion disk radiation are nearly always made with respect to
models (such as the ``multi-color disk": Ebisawa et al. 1994, Shimura
\& Takahara 1995) that assume surface brightnesses following the
Novikov-Thorne prescription or its Newtonian simplification.  As we
have shown here, this is unlikely to be a good description of the
radiation edge.  There is then little basis for the
frequently-performed further step of using these model fits, in which
the radiation edge is assumed to lie at exactly $r_{ms}$ to infer
a black hole's mass and spin.

\acknowledgments

This work was supported by NASA grant NAG5-9187 to JHK, and NSF grant
AST-0070979 and NASA grant NAG5-9266 to JFH.
Computational support was provided by the San Diego Supercomputer
Center of the National Partnership for Advanced Computational
Infrastructure, funded by the NSF, and by the Legion project at the
University of Virginia.

\end{document}